\documentclass{article}

\usepackage{authblk}
\usepackage{graphicx}
\usepackage{hyperref}

\usepackage{amsmath} 
\usepackage{bm}
\usepackage{float}

\usepackage{geometry}
\newcommand{\mitbf}[1] {\hbox{\mathversion{bold}${#1}$}}
\newcommand{\curl}{\nabla\times}
\newcommand{\di}{\nabla\cdot}
\newcommand{\grad}{\nabla}
\newcommand{\pdt}[1]{\frac{\partial #1}{\partial t}}

\newcommand{\B}{\mitbf{B}}
\newcommand{\vv}{\mitbf{v}}
\newcommand{\J}{\mitbf{J}}
\newcommand{\E}{\mitbf{E}}
\newcommand{\rr}{\mitbf{r}}
\newcommand{\vz}{\mitbf{0}}
\newcommand{\er}{\mitbf{e}_r}
\newcommand{\BM}{\B_\mathrm{c}}
\newcommand{\Boi}{\B_\mathrm{oi}}
\newcommand{\Joi}{\J_\mathrm{oi}}
\newcommand{\Eoi}{\E_\mathrm{oi}}
\newcommand{\BoiT}{\B_\mathrm{oi}^{\mathrm{T}}}
\newcommand{\BoiP}{\B_\mathrm{oi}^{\mathrm{P}}}

\newcommand{\mo}{\mu_0}

\newcommand{\ondG}{\quad\mathrm{on}\;\partial G}
\newcommand{\inG}{\quad\mathrm{in}\;G}
\newcommand{\inA}{\quad\mathrm{in}\;A}
\newcommand{\jmx}{J}
\newcommand{\Bjm}[1]{B_{jm}^{(#1)}}
\newcommand{\Sjm}[1]{\mitbf{S}_{jm}^{(#1)}}

\newcommand{\Yjm}{Y_{jm}}

\newcommand{\Gjm}{G_{jm}}
\newcommand{\kmx}{P}
\newcommand{\dV}{\mathrm{d}V}

\begin{document}

\title{Satellite monitoring of long period ocean-induced\\ magnetic field variations}

\author[1,*]{C. C. Finlay}
\author[2]{J.  Vel\'{i}msk\'{y}}
\author[1]{C. Kloss}
\author[1]{R. M. Blangsb{\o}ll}

\affil[1]{Department of Space Research and Technology, Technical University of Denmark, Lyngby, Denmark}
\affil[2]{Department of Geophysics, Faculty of Mathematics and Physics, Charles University, Prague, Czech Republic}



\affil[*]{e-mail: cfinl@dtu.dk}

\maketitle

\begin{abstract}
Satellite magnetic field observations have the potential to provide valuable information on dynamics, heat content and salinity throughout the ocean.  Here we present the expected spatio-temporal characteristics of the ocean-induced magnetic field at satellite altitude on periods of months to decades.  We compare these to the characteristics of other sources of Earth's magnetic field, and discuss whether it is feasible for the ocean-induced magnetic field to be retrieved and routinely monitored from space. 

We focus on large length scales (spherical harmonic degrees up to 30) and periods from one month up to five years.  To characterize the expected ocean signal we make use of advanced numerical simulations taking high resolution oceanographic inputs and solve the magnetic induction equation in 3D including galvanic coupling and self induction effects. We find the time-varying ocean-induced signal dominates over the primary source of the internal field, the core dynamo, at high spherical harmonic degree with the cross-over taking place at degree 13 to 19 depending on the considered period.  The ionospheric and magnetospheric fields (including their Earth induced counterparts) have most power on periods shorter than one month and are expected to be mostly zonal in magnetic coordinates at satellite altitude.  

Based on these findings we discuss future prospects for isolating and monitoring long period ocean induced magnetic field variations using data collected by present and upcoming magnetic survey satellites.

\end{abstract}



Satellite observations are essential to our modern view of the ocean.  They provide global information on circulation patterns \cite{Wunsch1998, Knudsen2021}, temperature \cite{Merchant2019}, salinity \cite{Boutin2021}, and biochemistry such as chlorophyll \cite{Hu2012} on horizontal scales ranging from 10 to 10,000 km and spanning timescales of weeks to decades, enabling variability and trends in ocean properties to be monitored.  These observations are nevertheless heavily focused on the ocean surface; a full understanding of the ocean and its changes also requires information on processes happening at depth.  Float programs, in particular ARGO, have in recent years provided much valuable information in this regard \cite{RoemmichGilson2009}.  Our knowledge of ocean flows, temperature and salinity as a function of depth nonetheless remains incomplete, especially at depths below 2000\,m.

Satellite measurements of gravity and magnetic fields involve an intrinsic integration of ocean properties over depth, so they provide complementary observational constraints on the ocean as a whole.  For example, gravity measurements provide constraints on total ocean mass variations and ocean bottom pressure variations which depend on gravity via density integrated over a vertical water column \cite{Hughes2018}.  The GRACE mission has already provided extremely valuable information in this regard, and the upcoming MAGIC mission aims to push such monitoring to higher spatial and temporal resolution \cite{Daras2024}.  Our focus here is instead on satellite measurements of the magnetic field and the information they can provide on integrated ocean properties on large length scales (of order 1000 km and larger) and long (monthly to decadal) time scales.  Magnetic fields are generated by motional induction in the ocean, when electrically-conducting water moves across magnetic field lines originating in Earth's core \cite{Sanford1971,Tyler1995, Minami2017}.  We refer here to such fields as \textit{Ocean-Induced Magnetic Fields} (OIMF), though it is important to note that they also depend on the electrical conductivity of the solid Earth where the induced currents close.

Two classes of oceanic motions contribute to the OIMF. First, and so far best studied, are the magnetic signals due to ocean motions driven by periodic gravitational forcing of the luni-solar tides. Such tidal flows are predominantly two-dimensional, and they give rise to pronounced magnetic signals at the discrete frequencies of the individual tidal constituents \cite{Tyler2003, Saynisch2021,Grayver2024}, although non-linear
compound tides also exist \cite{Einspigel2017}.  The second class, which is our focus here, involves magnetic signals resulting from the  general circulation of the ocean and its dynamics.  The latter motions are driven both by winds and buoyancy (also known as thermohaline) effects including surface temperature variations and freshwater release, with timescales ranging from days up to centuries\cite{Talley2011,WilliamsFollows2011,Wunsch2015}.  These general  circulation motions of the ocean are again largely horizontal, although the small vertical transport plays a crucial role in the dynamics of the thermohaline part of the circulation. 

The kinetic energy of near-surface motions is dominated by mesoscale eddies (on length scales of hundreds of kms and shorter, and with typical timescales of days to months) \cite{Chelton2011}, however, it is larger (basin-scale) gyres and overturning circulations that dominate thermohaline transport processes over longer timescales. Examples of large-scale ocean currents systems relevant here include the Antarctic Circumpolar Current (ACC), the Atlantic Meridional Overturning Circulation (AMOC) including the Gulf stream, the Indian Ocean Circulation with its Monsoon dynamics,  the Pacific Equatorial Currents that are related to the El Ni\~no-Southern Oscillation climate pattern, and the North and South Pacific Subpolar Currents including the Kuroshio Current  \cite{Steele2009,Talley2011} .  The evolution of these large-scale current systems and their depth variations are of great interest for understanding Earth's climate system \cite{WilliamsFollows2011}. The OIMF is expected to carry volume-integrated information on such processes, if it can be reliably extracted.

In addition to being dependent on the amplitude and geometry of flows in the ocean, the OIMF also depends on the electrical conductivity of sea water, which in turn depends on temperature and salinity
\cite{Millero2010,Grayver2021}. The ocean's electrical conductivity thus
varies with geographic location, depth and time, including seasonal variations and long-term trends.  Variations in the large-scale OIMF will therefore also be produced by large-scale changes in ocean temperature (i.e. heat content) and salinity.  

Over the past decade there has been significant progress in understanding and simulating the OIMF.  There is now consensus regarding the basic features and expected amplitude of the time-averaged ocean-induced magnetic field based on advanced simulations that make use of oceanographic data \cite{Sachl2019}.  Simulations suggest the OIMF originates primarily from the upper 2000m of the ocean where motions are larger and the temperature (hence electrical conductivity) is higher \cite{Irrgang2018}.  At Earth's surface and satellite amplitude it is dominated by the signature of the Antarctic Circumpolar current \cite{Stephenson1992,Vivier2004,Glazman2005,Manoj2006}. There is however less consensus regarding the magnetic signal of ocean variability on timescales of months to decades; this depends on the input oceanographic information.  For such simulations to be accurate it is important to treat ocean motions, temperature and salinity (as well as the derived electrical conductivity) in a self-consistent manner.  This requires input of realistic 4D (3D plus temporal) ocean properties while acknowledging that even the most up-to-date oceanographic information is incomplete.  Systematic numerical experiments are required to address such uncertainties; here we present only an illustrative example of the OIMF from one such simulation, a more extensive set of simulations is underway and will be reported in a future study. 

Turning to observations of the OIMF, ocean tidal signals can now be routinely extracted from satellite magnetic measurements. The largest and easiest to extract is the $M_2$ tide \cite{Tyler2003, Sabaka2016, Sabaka2020} which is available as an L2 Data Product (SW$\_$MTI$\_$SHA$\_$2C) from ESA's \textit{Swarm} satellite mission.  The magnetic signals of the $N_2$ and $O_1$ tides have also been extracted \cite{Grayver2019, Saynisch2021} and most recently the weaker $Q_1$ tide (amplitude at satellite altitude of order 0.1 nT) has been convincingly isolated \cite{Grayver2024}. In the most recent study by Grayver et al. \cite{Grayver2024} spherical harmonic models of the internal magnetic field with a specific period are fit to observations from the CHAMP and \textit{Swarm} satellites, using only night side data and after correcting for core and magnetospheric fields.  The observed OIMF tidal signals agree well with predictions from numerical simulations taking the rather well known tidal flows as input.  These observations of the OIMF tidal signals are an important demonstration that the OIMF can be reliably retrieved from satellite magnetic measurements, despite the small signal amplitude, given appropriate processing and modelling of the signal. Trends in tidal properties, which may reflect trends in temperature, salinity or changes in underlying ocean circulation patterns, are of great interest; this is the next target for studies of the tidal OIMF.

Turning to the more general OIMF signal from the ocean general circulation, as far as we are aware there has only been one published claim to have detected this, based on observations from the CHAMP satellite \cite{Golubev2012}.  This result has however not yet been independently reproduced by other workers.  Recently Hornschild et al. \cite{Hornschild2022} sought to make progress by assuming the OIMF signal from a simulation was correct, except for a spatially-varying rescaling factor, which they estimated.  There is clearly still much to do regarding observational studies of the general circulation driven OIMF. 

The aim of this article is to document the space-time characteristics expected for the general circulation component of the OIMF, and to compare these to other (non-oceanic) magnetic field signals measured by satellites.  Based on these comparisons the feasibility of extracting the OIMF will be assessed, and ways to make progress in this direction proposed.  In section 2 we review the physics of the OIMF, describe how it can be modelled and present its space-time characteristics.  In section 3 we review the present status of satellite geomagnetism, give an overview of sources of Earth's magnetic field and describe the space-time characteristics of fields originating in the core, crust, ionosphere and magnetosphere.  In section 4 we discuss strategies for extracting the OIMF before finishing with discussion and conclusions in section 5.
 
\section{Ocean-induced magnetic field}
\subsection{Physics of motional induction}

 Motion of an electrical conductor through a magnetic field induces an electric field across the conductor. In the presence of a closed conducting path this results in an electrical current and an associated motionally induced magnetic field.  The oceans produce a weak magnetic field in this way as they move through the magnetic field generated in Earth's core.  

The Ocean-Induced Magnetic Field (OIMF) $\Boi$ observed, for example at a satellite at location $(\rr;t)=(r,\Omega;t)=(r,\vartheta,\varphi;t)$ is an integral property of the electrical current density $\Joi$ induced in the oceans, including its closure in the solid Earth.  The Biot–Savart law describes this relationship and may be written
\begin{equation}
\Boi(\rr;t)= {\mu_0 \over 4 \pi} \int\limits_{G} {\Joi(\rr',t) \times (\rr- \rr') \over \left| \rr- \rr' \right|^3} \, \dV' 
\end{equation}
where $G = G_\mathrm{o} \cup G_\mathrm{se}$, is the volume of the oceans and the solid Earth, taken here to be the region $r\le a$ where $a$ is the mean spherical radius of Earth's surface and $\mo=\mathrm{const}$ is the magnetic permeability of free space. 

The current density $\Joi(\rr;t)$ is specified by Ohm's law for a moving conductor to be 
\begin{equation}
\label{eq:Ohm}
\Joi=\sigma  \left[(\vv \times \B) + \Eoi \right].
\end{equation}
$\Joi(\rr;t)$ thus depends point-wise on the electrical conductivity $\sigma(\rr;t)$,  the ocean flow velocity $\vv(\rr;t)$, the total magnetic field $\B(\rr;t)$, and the electric field $\Eoi(\rr;t)$ induced by time changes of the magnetic field. $\B(\rr;t)=\BM(\rr;t)+\Boi(\rr;t)$ includes both a background magnetic field generated mostly in Earth's core $\BM(\rr;t)$ and the OIMF $\Boi(\rr;t)$.

The electric field $\Eoi$, current density $\Joi$ and magnetic field $\Boi$ associated with motional induction in the oceans are related by Maxwell's equations. Under the quasi-static approximation that is appropriate for the slow changes considered here, these take the form
\begin{equation}
\label{eq:Maxwell}
\nabla \times \Eoi = - {\partial \Boi \over \partial t}  \qquad \mbox{and} \qquad   \nabla \times \Boi = \mu_0 \Joi.
\end{equation}

Combining (\ref{eq:Ohm}) and (\ref{eq:Maxwell}) and making use of the fact that $|\BM| \gg |\Boi|$, gives 
\vspace{-0.1cm}
\begin{equation}
{\partial \Boi \over \partial t} = \nabla \times (\vv \times \BM) -{1 \over \mu_0}\nabla \times \left[{1 \over \sigma} (\nabla \times \Boi) \right]. \nonumber
\end{equation}
Known in magnetohydrodynamics as the magnetic induction equation, this specifies the time evolution of $\Boi$ given the ocean flow velocity $\vv(\rr;t)$,  the electrical conductivity of the oceans and solid Earth $\sigma(\rr;t)$, and the core field $\BM(\rr;t)$. 

Taking into account the divergence free condition that holds for all magnetic fields, and an insulating boundary condition at the Earth's surface, we therefore have the following set of equations that can be used to simulate the ocean-induced magnetic field,
\begin{eqnarray}
\label{eq:EMI}
{1 \over \mu_0}\curl\left(\frac{1}{\sigma}\curl\Boi\right) + \pdt{\Boi} & = &\curl\left(\vv\times\BM\right) \inG, \\
\di\Boi & = & 0 \inG, \\
\label{eq:BC}
\Boi & = & -\grad U \ondG,\\
\label{eq:Lap}
\nabla^2 U & = & 0 \inA,\\
\label{eq:UBC}
U & = & \mathcal{O}\left(r^{-2}\right)\,\mathrm{as}\,r\rightarrow\infty,\\
\label{eq:init}
\left.\Boi\right|_{t=0} & = & \B_0,
\end{eqnarray}
where $A$ is the insulating atmosphere $r\ge a$, $\partial G$ is the Earth/atmosphere interface, $U(\rr;t)$ is the magnetic scalar potential in $A$, which decreases with
radial distance in the absence of external sources, and $\B_0(\rr)=\Boi(\rr;t=0)$ is the initial condition for the ocean-induced magnetic field in $G$.

The only physical approximation involved in the above equations is the quasi-static approximation of Maxwell's equations, where displacement currents are neglected since low frequencies are considered. Time-dependent electrical conductivities, reflecting changes in the salinity and temperature of the oceans, are permitted. The time derivative of $\Boi$ in the induction equation
(\ref{eq:EMI}) leads to inductive coupling between the highly conductive oceans, the moderately conductive seafloor sediments, and the poorly conductive lithosphere and mantle, as well as self-induction.
Spatial variations of electrical conductivity imply the presence of galvanic coupling, as the electric currents prefer the least-resistive path through the heterogeneous medium.
This can be demonstrated by writing separately the poloidal and toroidal parts of equation (\ref{eq:EMI}),
\begin{eqnarray}
\label{eq:EMIP}
\left[{1 \over \mu_0}\curl\left(\frac{1}{\sigma}\curl\BoiP\right)\right]_\mathrm{P} + \pdt{\BoiP} & = &\left[\curl\left(\vv\times\BM\right)\right]_\mathrm{P}
- \left[{1 \over \mu_0}\curl\left(\frac{1}{\sigma}\curl\BoiT\right)\right]_\mathrm{P}, \\
\label{eq:EMIT}
\left[{1 \over \mu_0}\curl\left(\frac{1}{\sigma}\curl\BoiT\right)\right]_\mathrm{T} + \pdt{\BoiT} & = &\left[\curl\left(\vv\times\BM\right)\right]_\mathrm{T}
- \left[{1 \over \mu_0}\curl\left(\frac{1}{\sigma}\curl\BoiP\right)\right]_\mathrm{T}.
\end{eqnarray}
The poloidal and toroidal magnetic fields, and corresponding
toroidal and poloidal electric currents, are generated respectively by the toroidal and poloidal components of the Lorentz force per unit charge, $\vv\times\BM$ \cite{Velimsky2019}.
{\v{S}}achl et al. \cite{Sachl2019} have demonstrated that it is important to include galvanic coupling effects when simulating the OIMF by simultaneously modelling both the toroidal and poloidal fields. Equations \ref{eq:EMIP} and \ref{eq:EMIT} show that lateral variations of electrical conductivity couples the poloidal and toroidal fields, and that energy is exchanged between them.

\subsection{Numerical simulations}
Below we present results from numerical simulations of the OIMF  based on a time-domain, spherical harmonic-finite element solver of equations (\ref{eq:EMI})-(\ref{eq:init}).  The numerical scheme was
originally developed to model the electromagnetic response of an electrically heterogeneous Earth to time variations of the magnetic field of external (i.e., magnetospheric
or ionspheric) origin \cite{Velimsky2005,Velimsky2013,Velimsky2021b}, and later modified to simulate the OIMF by implementation of an internal forcing
due to ocean flows \cite{Sachl2019,Velimsky2019,Velimsky2021a}.

The magnetic field in $G$, and the scalar magnetic potential in $A$ are respectively expanded into truncated series of vector and scalar spherical harmonics, 
\begin{eqnarray}
\Boi(\rr;t)& = & \sum\limits_{j=1}^{\jmx}\sum\limits_{m=-j}^{j}\sum\limits_{\lambda=-1}^{+1}
\Bjm{\lambda}(r;t)\Sjm{\lambda}(\Omega), \\
U(\rr;t) & = & a \sum\limits_{j=1}^{\jmx}\sum\limits_{m=-j}^{j}
\Gjm(t)\left(\frac{a}{r}\right)^{j+1}\Yjm(\Omega).
\end{eqnarray}
Here
\begin{equation}
    \Yjm(\Omega)= \sqrt{\frac{2j+1}{4\pi}\frac{(l-|m|)!}{(l+|m|)!}}P_l^{|m|}(\cos\vartheta)\left\{\begin{array}{lcl}
    (-1)^m\sqrt{2} \sin(|m|\varphi) & \mathrm{if} & m<0,\\
    1 & \mathrm{if} & m=0,\\
    (-1)^m\sqrt{2} \cos(|m|\varphi) & \mathrm{if} & m>0
    \end{array}\right.
\end{equation}
are the real, fully normalized scalar spherical harmonics, and the poloidal vertical, toroidal, and poloidal horizontal vector
spherical harmonics are defined as
\begin{eqnarray}
\Sjm{-1}(\Omega)& =& \er\,\Yjm(\Omega),\\
\Sjm{0}(\Omega)& =& \er\times\nabla_\Omega\Yjm(\Omega),\\
\Sjm{+1}(\Omega)& =& \nabla_\Omega\Yjm(\Omega),
\end{eqnarray}
respectively. The magnetic field is further parameterized in the radial direction using piecewise-linear finite elements
over the radial coordinate discretized by $0=r_1 < r_2 < \dots < r_{\kmx} < r_{\kmx+1}=a$,
\begin{eqnarray}
\Bjm{\lambda}(r;t) & = & \sum\limits_{k=1}^{\kmx+1} \Bjm{\lambda,k}(t)\psi_k(r),\\
\psi_k(r)& = & \left\{\begin{array}{lcl}
{r-r_{k-1} \over r_{k}-r_{k-1}} & \mathrm{for} & r\in\left[r_{k-1},r_k\right],\\
{r_{k+1}-r \over r_{k+1}-r_k} & \mathrm{for} & r\in\left[r_k,r_{k+1}\right],\\
0 & \mathrm{elsewhere.} &
\end{array}\right.
\end{eqnarray}
An integral formulation of (\ref{eq:EMI}), a Galerkin scheme applied over the spatial coordinates,
and a Crank-Nicolson time-integration scheme
are then used to construct a linear system of equations. The initial condition $\B_0$ is chosen to be the solution of the stationary limit
of equation (\ref{eq:EMI}), found by setting $\pdt{\Boi}(\rr;0)=\vz$; this is equivalent to the homogeneous Neumann initial condition.  The linear system is assembled and solved using MPI parallelization in the time coordinate, where each node is assigned a contiguous block of time levels, and a full time series of $\Boi(\rr;t)$ is obtained by BiCGStab(2) matrix-free iterations. BiCGStab(2) is a Biconjugate Gradient Stablized method where a generalized minimal residual step follows every two biconjugate gradient steps. The implementation we employ here, known as \texttt{elmgTD\_mpi}, uses a fast matrix-vector product based on Gauss-Legendre quadrature and FFT in lateral coordinates, and a preconditioner derived from a simplified problem with radially
symmetric conductivity $\sigma_0(r)$ to accelerate the convergence of the iterations.

The simulation of the OIMF presented below is based on daily-mean velocities from the ocean state estimate
\texttt{ECCOv4r4} \cite{Forget2015, Fukumori2021}, which is the latest version of the Estimating the Circulation and Climate of the Ocean framework.  Here, we consider only the time interval 2013 to 2017 that overlaps with the lifetime of \textit{Swarm} satellite mission that is  presently surveying the geomagnetic field.  \texttt{ECCOv4r4} seeks to adequately fit global-scale observational constraints on the ocean state \cite{Fukumori2021} (including sea level, temperature profiles, salinity profiles, sea surface temperature, sea surface salinity, sea ice concentration, ocean bottom pressure, mean dynamic topography), while being a solution to a suitably adjusted oceanic general circulation model, the MITgcm \cite{Marshall1997, Adcroft2004}. More precisely, the adjoint of the MITgcm is used to estimate time-varying atmospheric forcing fields, the model's initial conditions in January 1992 and the model's time-invariant ocean mixing parameters.  These are then used in a forward run of the MITgcm from 1992 to 2017 which maintains dynamical and kinematic consistency with the equation of motion and conservation equations.

The three components of $\vv(\rr;t)$ from \texttt{ECCOv4r4} are interpolated laterally to the \texttt{elmgTD\_mpi}
computation grid of $180\times360$ points, which uses the Gauss-Legendre quadrature nodes in colatitude, and a regular step
in longitude. The ocean radial discretization of 50 layers, with layer thicknesses ranging from 10 m close to the ocean surface to
456.5 m in the deepest layer, is carried over to the \texttt{elmgTD\_mpi} solver. Below 6134.5 m, which is the
maximum depth of \texttt{ECCOv4r4}, zero velocities are prescribed and no forcing is present. The core 
field $\BM(\rr;t)$ is taken from the
spherical harmonic-based IGRF-13 model \cite{Alken2021}, including its linear secular variation, evaluated on the same grid as the ocean velocity.

The electrical conductivity model $\sigma(\rr;t)$ is assembled from three sources. In the oceans, the objectively
analyzed conductivity climatology from collocated salinity and temperature measurements of the World Ocean Atlas (WOA13,
\cite{Tyler2017}) is interpolated to the computation grid using bi-linear interpolation in lateral coordinates
and conductance-preserving vertical interpolation. Temporal variations of the ocean conductivity are taken into account by using the
WOA13 conductivity monthly means.  We acknowledge that the climatological conductivities used here are not fully consistent with the dynamics of the \texttt{ECCOv4r4} ocean flows; in future work we plan to compute ocean conductivities more directly from the ocean model's temperature and salinity via an equation of state. Immediately below the seafloor or below the continental surface, a simplified sediment model based
on assigned conductivity values for ocean, coastal, and continental sediments, and maps of their respective thicknesses 
is used \cite{Everett2003}. Again, bi-linear lateral and conductance-preserving vertical interpolation is applied.
Finally, below a depth of 26 km, the electrical conductivity follows a 1-D profile \texttt{MIN\_1DM\_0501},
which is constrained by joint inversion of magnetospheric and tidal signals observed by the \textit{Swarm} satellites  \cite{Grayver2017}.
In total, the oceans are spanned by 50 layers with 4-D spatial and temporal conductivity variations, on a grid consistent with that used for the $\vv\times\BM$ forcing term, while the sediments are discretized by 40 layers with spatially 3-D, time-invariant
conductivity, and the mantle and core cover 116 1-D layers, with thickness gradually increasing with depth.

\subsection{Global diagnostics}
Below we make use of the following global diagnostics to characterize the typical  spatial and temporal scales of the simulated OIMF, and compare these to other expected geomagnetic signals, which are presented in detail below. 

To document the power in the magnetic field as a function of spatial lengthscale, we employ the Lowes-Mauersberger spherical harmonic power spectra $R_j$ \cite{Mauersberger1956, Lowes1966}.  This is the mean squared vector magnetic field associated with spherical harmonic degree $j$, which can be computed as
\begin{equation}
R_j = \frac{(j+1)(2j+1)}{4\pi} \left( {a \over r}\right)^{(2j+4)} \sum\limits_{m=-j}^j |\Gjm|^2.
\end{equation}
where $\Gjm$ are the fully normalized spherical harmonic coefficients of the scalar magnetic potential $U$.

To characterize the temporal distribution of power as a function of frequency we consider the median power spectral density $\hat{P}_{B_r}$ of the radial component of the magnetic field $B_r(\rr;t)$, evaluated on an approximately equal area grid at the radius of interest.  We compute the power spectral density using Welch's method \cite{Welch1967}, and define $\hat{P}_{B_r}$ such that  
\begin{equation}
\hat{P}_{B_r} (f_n) = \textit{Med}\left\{ {L \over Q} \left[ {1 \over L} \sum\limits_{q=0}^{L-1} B_r(t_q) W(q) e^{-2iqn/L} \right]^2 \right\}  
\end{equation}
where the input series at each gridpoint is $B_r(t_q)$, $f_n$ are the discrete frequencies at which the power spectra density is estimated, $L$ is chosen to be the length of the time series since we are interested in periods comparable to that of the full the time series. $W(q)=\sin^2(\pi q/L)$ is a Hann window weighting function,  $Q= 1 /L \sum_q W^2(q)$ and $i=\sqrt{-1}$. $\textit{Med}$ denotes the median of the power spectral densities computed at each gridpoint on the spherical surface.  When the signal considered is the OIMF, a mask is employed so only grid points within or above the ocean are considered. 


\subsection{Filtering}
To study the OIMF signal, and other geomagnetic signals, at different spatial scales we truncate the spherical harmonic expansion at chosen minimum or maximum degrees $J$.  For $J>>1$ the lengthscale associated with spherical harmonic degree $J$ on a sphere of radius $r$ is approximately $2\pi r/J$.

We investigate different temporal scales filtering in time so as to highlight particular period bands of interest.  We first remove the mean value and then use a forward-backward implementation of a 2nd order Butterworth filter (so the combined action of the filter is 4th order and the filter is by construction zero phase).  Inital conditions are set following the method of Gustafsson \cite{Gustafsson1996} in order to minimize warm-up transients.

\subsection{Simulated Ocean-Induced Magnetic Field}
In this section we present the global characteristics of the general circulation OIMF signal simulated using the method described above. Results are presented at an altitude of 450\,km, a typical altitude of the \textit{Swarm} magnetic survey satellites, which since 2013 have provided high quality global measurements of the geomagnetic field (see Sect. \ref{Sect:Sat_geomag} for more details).  We focus here on the radial field component.

Fig.\,\ref{fig:Br_ocean_,map_satalt} presents maps of the radial component of the OIMF at 450\,km, simulated using the method described above, for the illustrative epoch of 2016.5. It shows the full,  unfiltered in time, signal (top panel), and the signal that is restricted to an annual period band, defined here to be 240 to 540 days (bottom panel).  At satellite altitude the amplitude of the basic OIMF signal is estimated to be of order 1\,nT while the annual signal has amplitude of order 0.1\,nT, consistent with previous estimates \cite{Manoj2006,Sachl2019}. 

\begin{figure}[!h]
\centerline{Full signal \qquad \qquad}
\centering\includegraphics[width=5.0in]{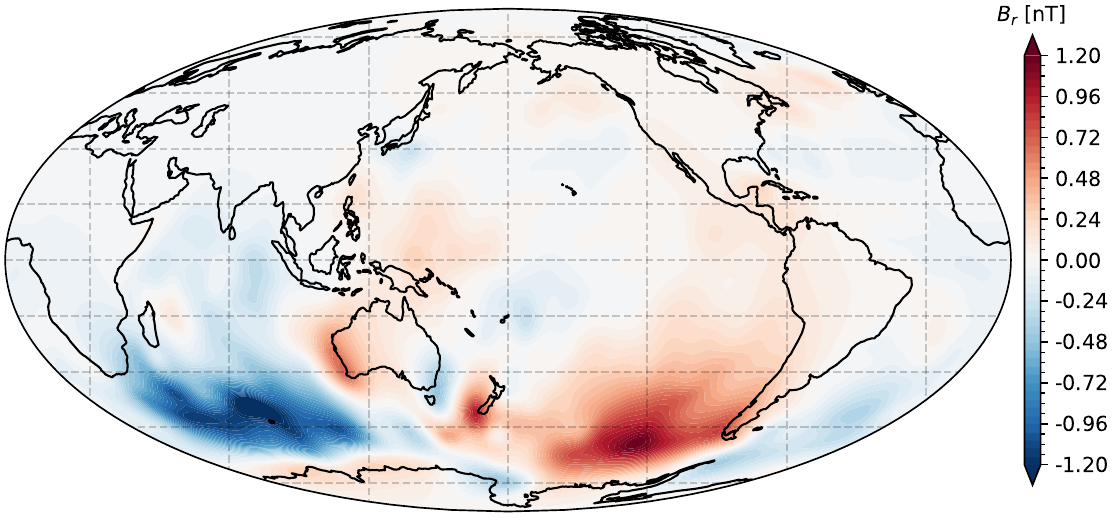}
\centerline{1 year period band \qquad }
\centering\includegraphics[width=5.0in]{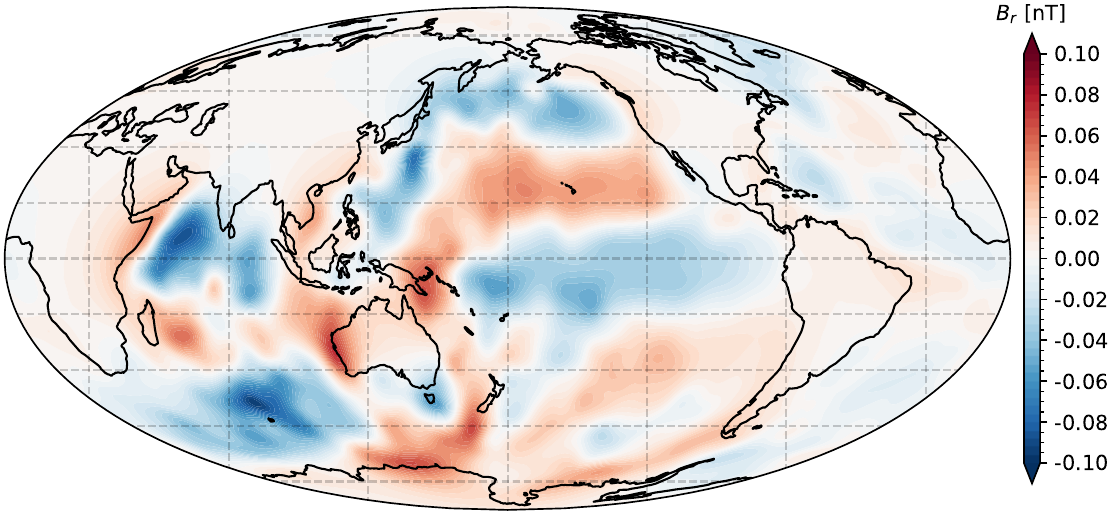}
\caption{Map of the radial component of the simulated ocean-induced magnetic field at satellite altitude, 450km, unfiltered in time (top) and filtered to retain a period band centered on 1 year (bottom), passband 240 to 540 days, in epoch 2016.5. The spherical harmonic expansion was truncated at degree 60.}
\label{fig:Br_ocean_,map_satalt}
\end{figure} 

\begin{figure}[!h]
\centering\includegraphics[width=5.0in]{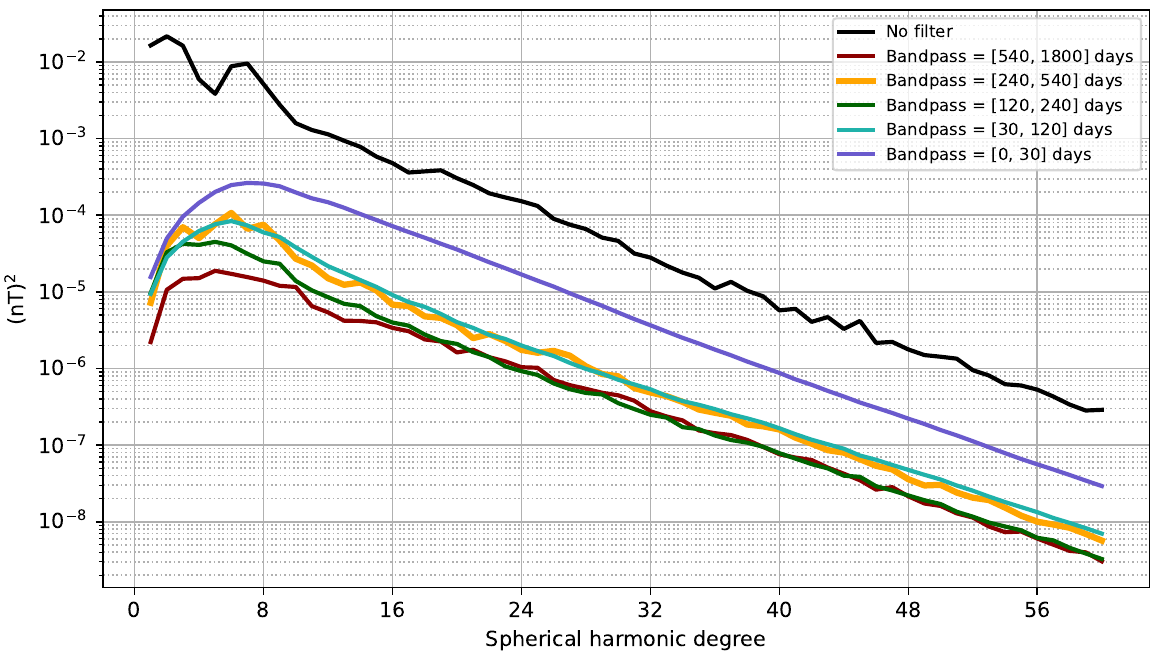}
\caption{Spatial spectra of the mean squared vector magnetic field of the simulated OIMF as a function of spherical harmonic degree, at altitude 450km, averaged over time interval 2013.0 to 2018.0.  For the unfiltered in time signal (black) and in interannual, annual, semi-annual, monthly and sub-monthly period bands (colours, see legend).}
\label{fig:spatial_spec_ocean_satalt}
\end{figure}

The unfiltered OIMF is most prominent in the Southern oceans, particularly in the Southern Pacific south-east of South America and in the Southern Indian ocean, both associated with the powerful Antarctic Circumpolar Current (ACC) system.  This first order structure is seen at all times and reflects the time-averaged ocean general circulation.  The spherical harmonic power spectra of the OIMF signal at satellite altitude, presented in Fig.\,\ref{fig:spatial_spec_ocean_satalt}, shows this ACC dominated time-averaged OIMF signal (black curve is the unfiltered OIMF) has peaks in the spherical harmonic power spectra at degrees 2-3 and 6-7. 

If one filters the OIMF signal in time so as to remove the time-average, signals from other parts of the ocean become more apparent.  For example, in Fig.\,\ref{fig:Br_ocean_,map_satalt} we also present a map in 2016.5 of the radial component of the OIMF at satellite altitude, but filtered to highlight a period passband around 1 year (240 to 540 days).  Here, weaker but  highly coherent signals are evident, particularly in the Pacific and Indian oceans. The signals in the Pacific have a clear east-west aligned structure and change sign with latitude;  three clear changes in the sign of $B_r$ are visible. Given their orientation, it seems likely that these are associated with seasonal variations in the Pacific Equatorial currents and the North and South Pacific currents that are located around latitudes 40 degrees north and south of the equator \cite{Talley2011}.  Similar  signals are also evident in $B_r$ over the Indian ocean.  Current systems in the Indian ocean are known to possess strong seasonal variations related to the Monsoon wind system\cite{Talley2011}.  There also seem to be some smaller-scale, meridionally oriented features, for example east of central Africa and south of Japan, that given their location may be related to   the Somali and the Kuroshio current systems.  

\begin{figure}[!h]
\centering\includegraphics[width=5.5in]{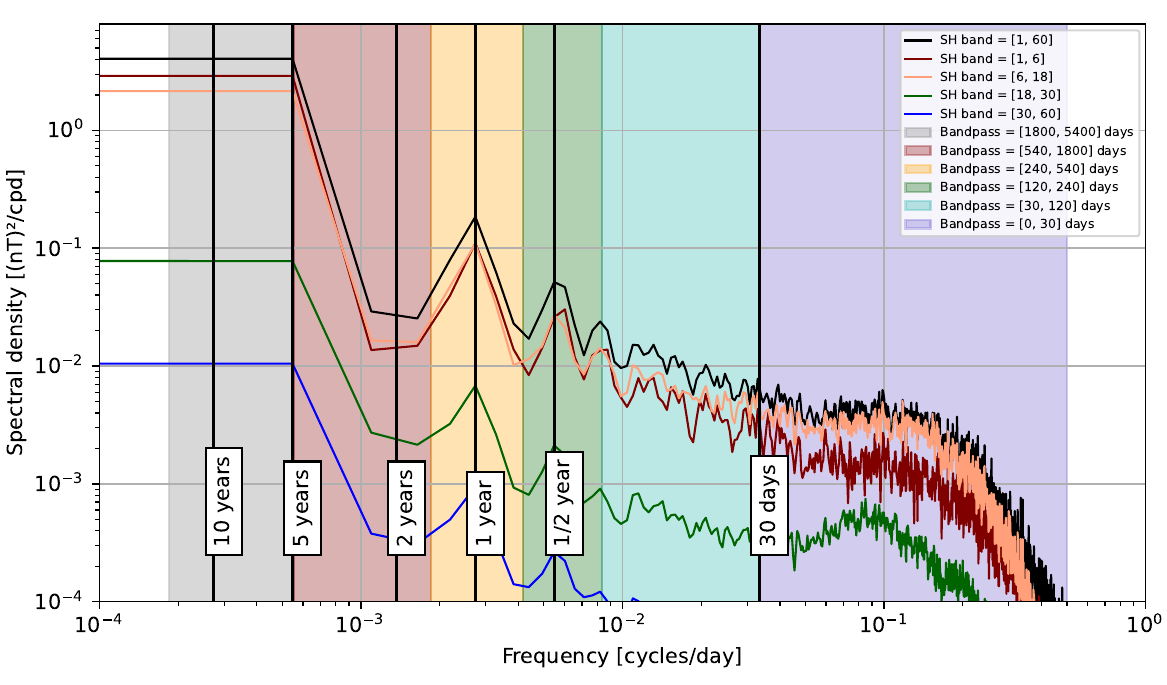}
\caption{Temporal power spectral density of the simulated radial component of the ocean-induced magnetic field at satellite altitude, 450km. Shown here is the median of power spectra computed on an approximately equal area grid of 3000 points, retaining only locations above the ocean (2095 points).  The power spectral density was computed using the Welch method, based on the simulated time interval from 2013.0 to 2018.0. Each solid line shows the spectra from a different range of spherical harmonics degrees, with the black line including all degrees up to 60.  The shaded regions indicate the period ranges considered in the spatial spectra. Note the use of log-log axis.}
\label{fig:temporal_spec_ocean_satalt}
\end{figure}

Spatial power spectra of the OIMF at satellite altitude, unfiltered (black curve) and filtered to highlight  interannual (540-1800 days), annual (540 to 240 days), semi-annual (240 to 120 days), monthly (120 to 30 days) and sub-monthly (less than 30 days) period bands,  are presented in Fig.\,\ref{fig:spatial_spec_ocean_satalt}. Aside from peaks present at degrees 2-3 and 6-7 in the unfiltered spectra, these generally show a power law decrease with increasing degree that reflects field sources located near to Earth's surface (recall ocean depths of up to 6\,km are much smaller than the 450\,km from Earth's surface to satellite altitude).  Re-plotting these spatial spectra at Earth's spherical reference radius (not shown) we find that they decrease in power much more slowly, dropping by about one order of magnitude between degree 5 and 60; the remaining slope is a consequence of the organization of electrical current sources below the ocean surface.  Periods shorter than 30 days are found to possess the majority of the variability in the simulated OIMF.  The peak power moves to higher spherical harmonic degree as the period shortens.  When interpreting these spatial spectra it should be remembered that they involve the temporal spectrum being integrated over period bands of physical significance that are however not uniformly spaced, see the coloured bands in Fig.\,\ref{fig:temporal_spec_ocean_satalt}; the spatial spectra for the shorter periods are also smoother as they effectively involve more complete time-averaging over the considered five years. 

Turning to the median temporal power spectral density of the radial component of the OIMF at satellite altitude, $\hat{P}_{B_r}(f_n)$ in Fig.\,\ref{fig:temporal_spec_ocean_satalt} we see that this shows most power per unit frequency at long periods with power decreasing as the period shortens down to 30 days; beyond this the power spectral density is flatter between 30 and 10 days before dropping off steeply below 10 days in this simulation.  A similar pattern is seen for all length scales (shown by the spectral densities for different spherical harmonic bands), except for that above degree 18 where we find the power increases for periods between 10 and 30 days, likely associated with mesoscale eddy signatures.  There are clear spectral peaks in the simulated OIMF at 1 and 0.5 yrs associated with annual and semi-annual variations.  When comparing the spatial and temporal spectra it should be remembered that our temporal power spectra densities have been plotted using a log scale in frequency, a consequence of this is that there are much fewer sampled frequencies at low periods.  When specific period bands are considered in the spatial spectra, the applied temporal filtering involves weighted sums in the frequency domain.  Since there are many more frequency bins to sum over for the bands at shorter periods, the overall signal in these bands is large, despite the power spectral density at these frequencies being lower.

Note that global averaging is inherent in all the above spectral diagnostics.  This may obscure interesting regional OIMF signatures.  Regional analysis of the OIMF signal can easily be performed using our analysis tools, but we postpone such  investigations to future studies.

\section{Satellite geomagnetism}
\label{Sect:Sat_geomag}
\subsection{Satellites dedicated to surveying Earth's magnetic field}
Having outlined the expected spatial and temporal characteristics of the OIMF at satellite altitude we now turn to the question of retrieving this signal from magnetic field measurements made by satellites dedicated to surveying Earth's magnetic field.  We begin by discussing the available measurements.

Beginning in the late 1960s, pioneering efforts in surveying Earth's magnetic field from space were made by NASA.  The POGO satellites provided the first global measurements of field intensity before the short-lived Magsat mission in 1979/1980 delivered the first vector field measurements from space. Since 1999 the past twenty five years have seen a series of satellite missions that have surveyed the Earth's magnetic field with increasing accuracy.  The Danish {\O}rsted satellite, launched in 1999, inaugurated this modern era of satellite geomagnetism.  It's vector field measurement system, with attitude information supplied by a non-magnetic star tracker co-located with a three-axis fluxgate magnetometer, was calibrated to absolute accuracy using an independent scalar magnetometer measuring the field intensity. The Argentine SAC-C satellite provided scalar intensity measurements from an {\O}rsted-like payload.  The German CHAMP satellite which flew from 2000 to 2010 built on the {\O}rsted satellite's measurement principles, but provided observations from lower altitudes of 310 to 450 km, closer to internal field sources including the ocean.  Since 2013 ESA's \textit{Swarm} satellite trio has been providing high quality measurements of the vector magnetic field with absolute accuracy.  Launched in 2013, \textit{Swarm} consists of three identical satellites,   a lower pair (\textit{Swarm} A and C) flying at altitudes 450-500\,km and an upper satellite (\textit{Swarm} B) typically about 50 km higher than the lower pair and drifting in local time.  The \textit{Swarm} mission has already provided a decade of high quality data for studying Earth's magnetic field. The Chinese CSES-1 satellite has provided useful scalar intensity measurement at mid and low latitudes since 2018.  Most recently the Macau Scientific Satellite (MSS-1), that like \textit{Swarm} measures the vector field with absolute accuracy, has been launched into a low inclination orbit that complements \textit{Swarm}'s polar orbits with a faster local time coverage at mid and low latitudes.  Taken together, we presently have twenty-five years of high quality measurements of Earth's magnetic field from low-Earth-orbit.  Table \ref{table_sats} summarizes the key characteristics of these magnetic survey satellites.

\begin{table}[!h]
\caption{Summary of satellite missions carrying absolute scalar magnetometers and used for geomagnetic field modelling.}
\label{table_sats}
\begin{tabular}{lllll}
\hline
Mission & Dates & Altitude (km) & Inclination (deg) & Magnetometer \\
\hline
POGO & 1965-1971 & 400-1500 & 82- 87 & Scalar Only \\
Magsat & 1979-1980 & 350-450 & 97 & Vector and Scalar\\
{\O}rsted & 1999-2013 & 650-850 & 97 & Vector and Scalar\\
CHAMP & 2000-2010 & 310-450 & 97 & Vector and Scalar\\
SAC-C & 2001-2004 & 698-705 & 97 & Scalar Only$^{*}$\\
\textit{Swarm} A, B \& C & 2013 - & 450-550 & 87, 88 \& 87 & Vector and Scalar\\
CSES-1 & 2018 - & $\sim$ 500 &  97 & Scalar Only$^{*}$\\
MSS-1 & 2023 - & $\sim$ 450 & 41 & Vector and Scalar\\

\hline 
\end{tabular}
\newline
\scriptsize{$*$=Payload included vector magnetometer, data not typically used for field modelling}
\vspace*{-4pt}
\end{table}

\subsection{An overview of geomagnetic field sources}
Measurements of Earth's magnetic field made by the low-Earth-orbit satellites described above consist of a superposition of signals from natural current sources present in the Earth's interior and in near-Earth geospace.  The largest source is powerful electrical currents generated by turbulent convection that acts as a dynamo in Earth's liquid metal outer core.  Close to Earth's surface, in the lithosphere, it is sufficiently cold for a small percentage of minerals to exist in a ferrimagnetic form which allows microscopic currents to be organized into large-scale magnetizations of geological structures that together give rise to the crustal or lithospheric field.  Also near to Earth's surface are the electrical currents induced by the motions of the ocean that are the primary focus here.  

Moving further out into the atmosphere, strong currents exist in the Ionospheric E-region around 110km altitude. High conductivity of the atmosphere due to ionization on the dayside and thermally and gravitationally driven motions (including solar and lunar tides) give rise to the Solar quiet (Sq) current system in this region.  The Sq current system moves with the sun and is modulated by seasonal variations in Earth's orbit and by solar activity.  In the high latitude ionosphere there are highly dynamic auroral electrojet current systems fed by field-aligned currents that originate primarily in the night-side of the outer magnetosphere (the magnetotail). These field aligned currents are also known as Magnetosphere-Ionosphere (MI) coupling currents.  Electrical conductivity is high in polar regions due to particles hitting the Earth's atmosphere there, being guided by Earth's magnetic field, as seen in auroral displays.  Significant currents also exist further out in the magnetosphere, in the radiation belts, in the magnetopause where the solar wind reaches Earth's magnetic field, and in the magnetotail on the nightside of the Earth.  The most important of these for satellite observations in low-Earth-orbit are the currents in the radiation belts, typically at distances of six to ten times Earth's radius, where circulating charged particles give rise to so-called magnetospheric ring currents.  These are dramatically enhanced when energy from solar disturbances is transferred into the magnetosphere following vigorous solar activity. Both magnetospheric and ionospheric currents vary rapidly and hence induce secondary currents in the electrically-conducting regions of the Earth, particular in the ocean, lithosphere and upper mantle.

\begin{figure}[!h]
\centering\includegraphics[width=4.0in]{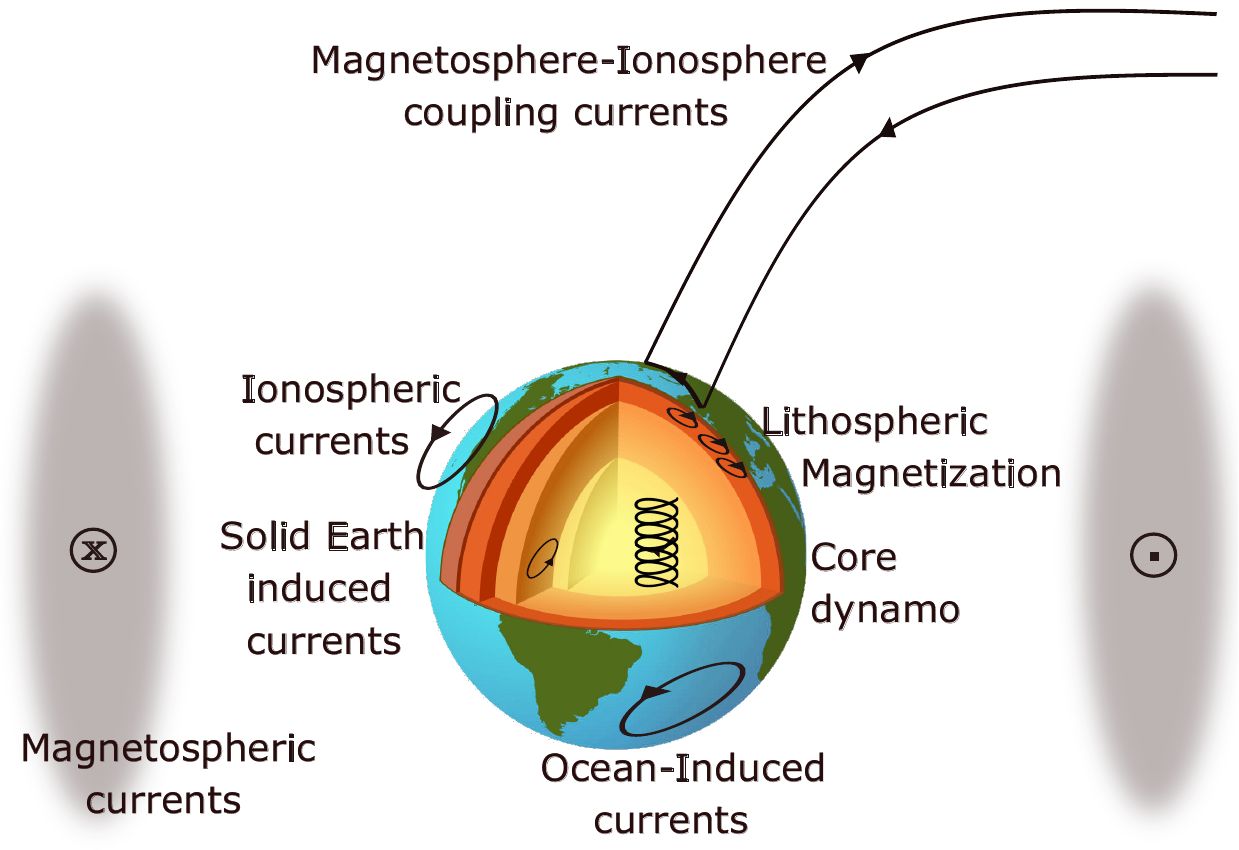}
\caption{A simplified cartoon depicting major near-Earth electrical current systems that contribute to Earth's magnetic field (not to scale, magnetospheric currents are in reality much more distant and with a complex spatio-temporal structure). 
}
\label{fig:current_systems}
\end{figure}

A simplified cartoon showing the basic geometry of these current systems is shown in Fig. \ref{fig:current_systems}. The real currents are of course much more spatially complex and not steady.  Table \ref{tab:sources} collects some summary information regarding these sources.  

\begin{table}[!h]
\caption{Summary of major sources of the near-Earth geomagnetic field. Information on expected length and time scales of the resulting fields at satellite altitude is shown via their spatial and temporal spectra in Figs.\, \ref{fig:spatial_spec_ocean_satalt}, \ref{fig:temporal_spec_ocean_satalt}, \ref{fig:core_spectra}, \ref{fig:iono_spectra},  \ref{fig:magneto_spectra}. }
\label{tab:sources}
\begin{tabular}{lll}
\hline
Source & Approx. geocentric & Order of magnitude$^*$  \\
 & radius of current system (km) & at satellite altitude (nT) \\
\hline
Core dynamo & 1220 - 3840 & 40,000 \\
Crustal/lithospheric magnetization & 6300 - 6371 & 10 \\
Ocean-induced currents & 6365-6371 & 1 \\
Solar-Quiet (Sq) ionospheric currents & $\sim$ 6480 & 20 \\
Auroral electrojet ionospheric currents & $\sim$ 6480 & 50 \\
Field-aligned MI coupling currents  & 6480- 100,000 & 100 \\
Magnetospheric ring current & 20,000-60,000 & 20 \\
Earth-induced currents & 3480-6371 & 10 \\
\hline 
\end{tabular}
\scriptsize{$*$=During geomagnetically quiet times of low solar activity, but note there can be considerably day-to-day variability for external sources.}
\vspace*{-4pt}
\end{table}

From Table \ref{tab:sources} the formidable challenge of extracting the ocean-induced magnetic field at satellite altitude is immediately apparent.  It is small compared to the signals from the other sources.  This means that careful selection, modelling, and filtering of the data based on the expected properties of both the OIMF and the those of the other sources, must be carried out.

An important aspect of the OIMF is that it is internal to the orbits of the satellites.  Using the tools of geomagnetic field modelling, and given good global data coverage, it is possible to separate sources internal and external to the observation altitude.  This means the major task in extracting the OIMF is to separate it from other internal sources: the core and the crustal/lithospheric fields, the ionospheric field (that is internal to satellites) and its associated Earth-induced fields, and Earth-induced currents driven by magnetospheric field variations.  In the sub-sections below we describe each of these in turn, presenting spatial and temporal characteristics from the state of the art models of each source with the aim of characterizing their properties at satellite altitude and understanding how they might be separated from the OIMF.

\subsection{Core and crustal fields}
\begin{figure}[!h]
\centering\includegraphics[width=3.15in]{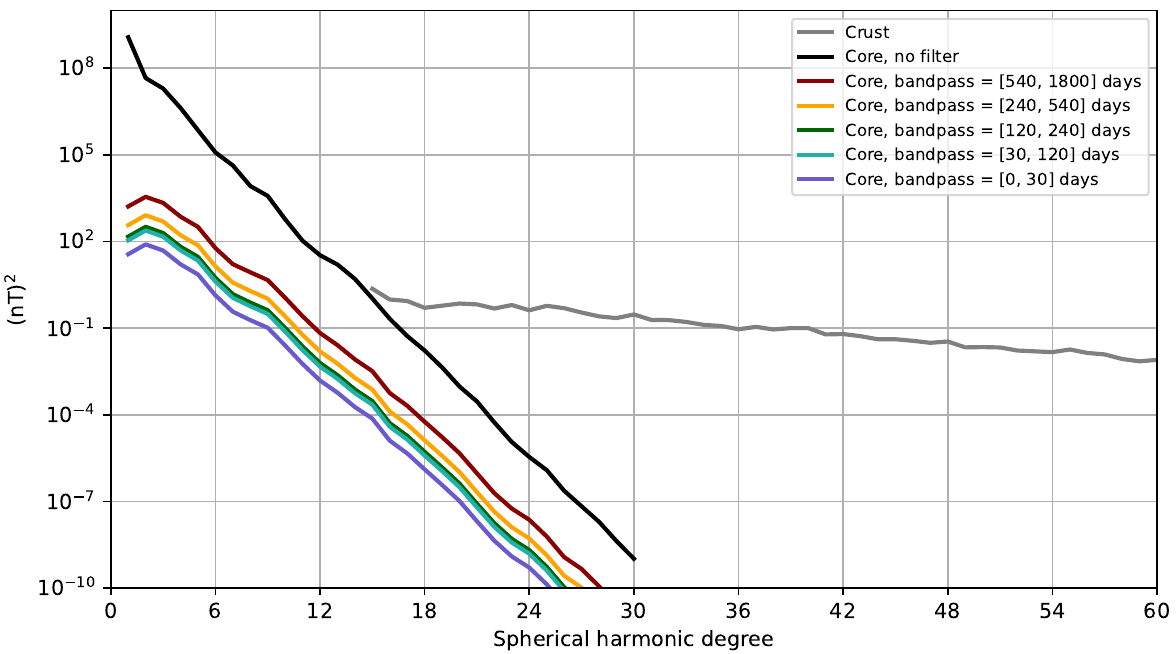}\hspace{0.05cm}\includegraphics[width=3.1in]{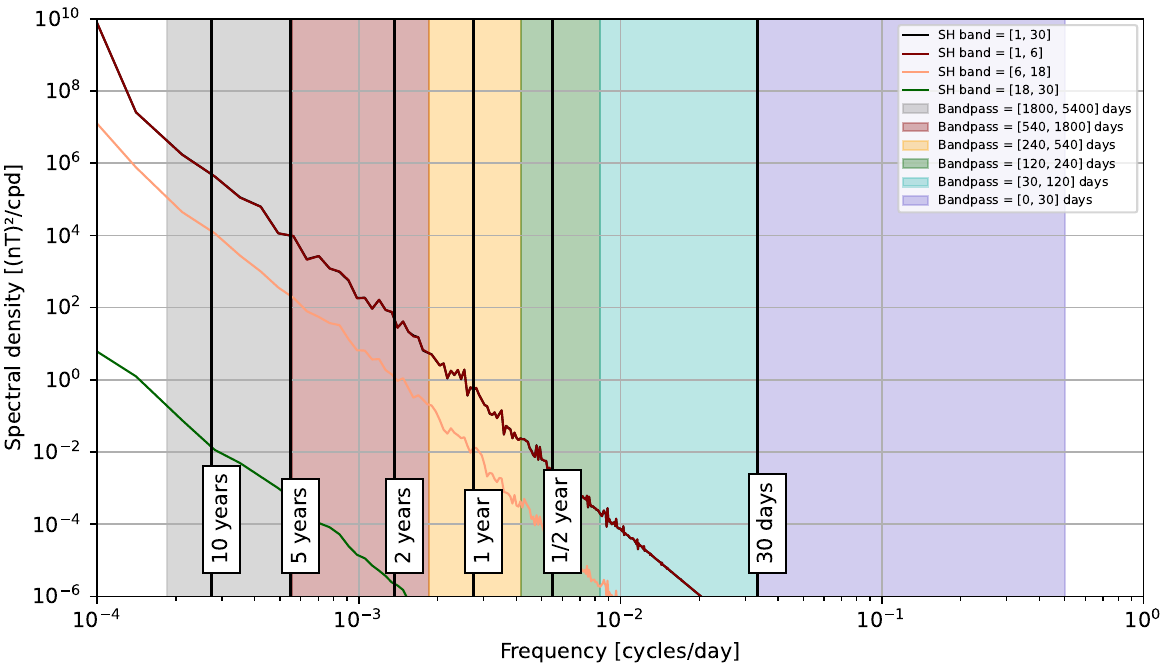}
\caption{Left: Spatial spectra of the mean squared vector magnetic field from a simulation of the core dynamo\cite{Aubert:2023} as a function of spherical harmonic degree, at satellite altitude (450\,km).  As well as the unfiltered signal (black curve), spatial spectra are shown for interannual, annual, semi-annual, monthly and sub-monthly period bands (coloured curves). The spatial power spectrum of the LCS-1 crustal field model\cite{Olsen:2017} is also shown (grey curve). Right: Temporal power spectral density $\hat{P}_{B_r}$ of the simulated radial component from the same core dynamo simulation at satellite altitude, unfiltered and in chosen spherical harmonic bands (curves can overlap). These curves are the median of power spectral densities, computed using a Welch-type method, of time series of $B_r$ computed on an approximately equal area grid of 3000 points at satellite altitude.}
\label{fig:core_spectra}
\end{figure}

As is clear from Table \ref{tab:sources}, the core dynamo is by far the largest source of magnetic fields measured close to the Earth.  In Fig.\ref{fig:core_spectra} we present time-averaged spatial spectra (left panel) and temporal spectra (right panel) of the core field at satellite altitude (450\,km altitude) from a recent Earth-like core dynamo simulation, the 100 percent path coupled-Earth dynamo \cite{Aubert:2023} whose construction involved the assimilation of geomagnetic observations.  The advantage of using a simulation here is that we obtain estimates of the core field's properties for length and time scales shorter than those traditionally considered for the core field, which are relevant for comparisons with the properties of the OIMF, while it also agrees well with traditional geomagnetic field models \cite{Alken2021} at larger length  scales.  Spatial spectra are also shown for the same period bands as considered in Fig.\,\ref{fig:spatial_spec_ocean_satalt}. The spatial spectra of the crustal (or more strictly the lithospheric) field is also included in Fig.\ref{fig:core_spectra}, taken from the LCS-1 \cite{Olsen:2017} model based on observations from the CHAMP and \textit{Swarm} satellite missions.

Although the core field is very strong on large length scales, its spatial spectra at satellite altitude drops off steeply as it is due to an internal source more than 3000km away.  This is the case for all the considered period bands.  The temporal power spectral density of the core field at satellite altitude also drops off steeply at all considered spatial wavelengths (up to spherical harmonic degree 30).   Compared with the shallower spatial and temporal spectra for the OIMF in Figs.\ref{fig:spatial_spec_ocean_satalt} and \ref{fig:temporal_spec_ocean_satalt} it is clear that at sufficiently short length and time scales one should expect the OIMF to dominate over the core field, we discuss this opportunity further in section \ref{sect:strategies}. 

Turning to the crustal field, like the OIMF this shows a shallower spatial spectrum at satellite altitude, since both are associated with sources close to Earth's surface.  However, the crustal field changes only very slowly in time \cite{Hulot2009}.  For the periods below 10 years considered here, time variations in the crustal field are expected to be small. So although it is very difficult to separate the steady component of the OIMF from the crustal field, the time-varying OIMF on periods of 10 years and shorter should not be significantly polluted by the crustal field.

\subsection{Ionospheric and associated Earth-Induced fields}
\begin{figure}[!h]
\centering\includegraphics[width=3.05in]{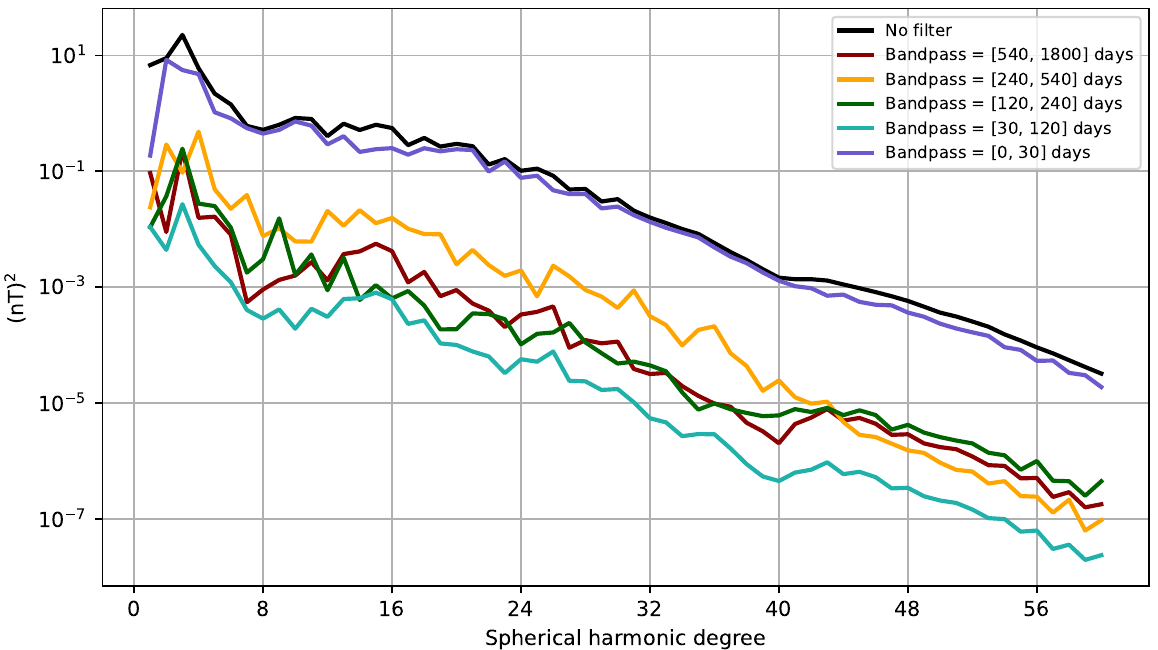} \hspace{0.05cm}\includegraphics[width=3.0in]{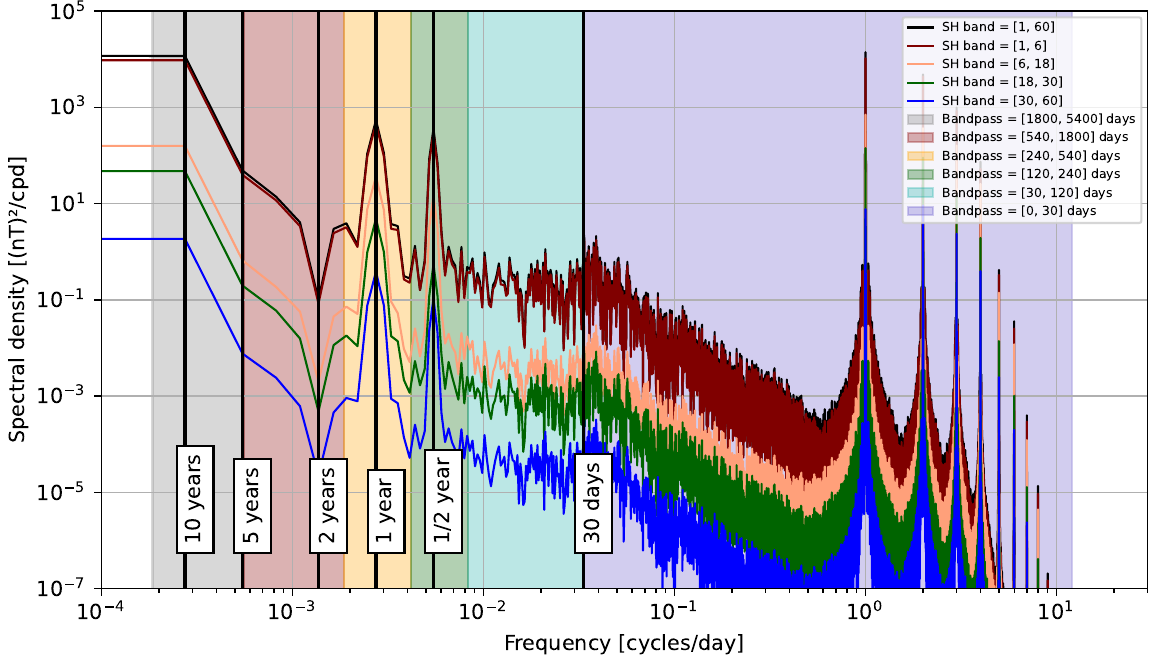}
\caption{Left: Spatial spectra of the mean squared vector magnetic field from the CI ionospheric field \cite{Sabaka:2018}, version 9, including its Earth-induced counterpart, as a function of spherical harmonic degree, at satellite altitude (450\,km).  As well as the unfiltered signal (black curve), spatial spectra are again shown for interannual, annual, semi-annual, monthly and sub-monthly period bands (coloured curves).  Right: Temporal power spectral density of the radial component of the ionospheric field and associated Earth-induced fields, again from the CI model, version 09, at satellite altitude. This is the median of the power spectral densities computed using a Welch-type method from time series of $B_r$ at 3000 approximately equal area grid points.}
\label{fig:iono_spectra}
\end{figure}

Fig.\ref{fig:iono_spectra} presents similar spatial and temporal spectra at satellite altitude from an observation-based model of ionospheric field and its associated Earth-induced counterparts. The model employed is the \textit{Swarm} comprehensive inversion \cite{Sabaka:2018}, version CI 09 which represents E-layer currents, with a specific focus on the Sq system, including explicit estimates of daily variations and related higher harmonics, annual and semi-annual variations, as well as solar modulation through a dependence on F10.7 solar flux.  The associated Earth-induced currents are calculated based on a 3D model of ocean and sediment electrical conductivity above a 1D conducting mantle \cite{Sabaka:2015}. 

The spatial spectra shown for different passbands in Fig.\ref{fig:iono_spectra} illustrate that the majority of the ionospheric and related induced signal occurs on timescales shorter than 30 days.  The temporal power spectral densities show distinct lines at one day and harmonics on all the investigated lengthscales.  Considering timescales longer than thirty days much of the ionospheric signal can therefore be filtered out.  

There are however significant ionospheric signals remaining in both the annual and semi-annual period bands; these are much larger than the annual and semi-annual signals seen in the temporal spectra for the simulated OIMF in Fig.\ref{fig:temporal_spec_ocean_satalt}.  Note that the spectra shown in Fig.\,\ref{fig:iono_spectra} are based on the modelled global ionospheric and induced fields each hour between 2014.0 and 2024.0, including both the day and night sides.  The amplitudes of the Sq currents will be weaker on the nightside;  we discuss these issues further in section \ref{sect:strategies}. 

We have not explicitly considered field-aligned magnetospheric-ionospheric coupling currents here.  Although of large amplitude these are expected to be of short length scale and to vary rapidly in time.  They are associated with toroidal magnetic fields, so should be orthogonal to poloidal fields of internal origin such as the OIMF.  They will however have an indirect impact through the E-layer currents they drive in the polar regions; although these are included, in principle, in the CI model we have employed here, its parameterization is not optimized for the polar region.

\subsection{Earth-Induced fields driven by Magnetospheric sources}

\begin{figure}[!h]
\centering\includegraphics[width=3.15in]{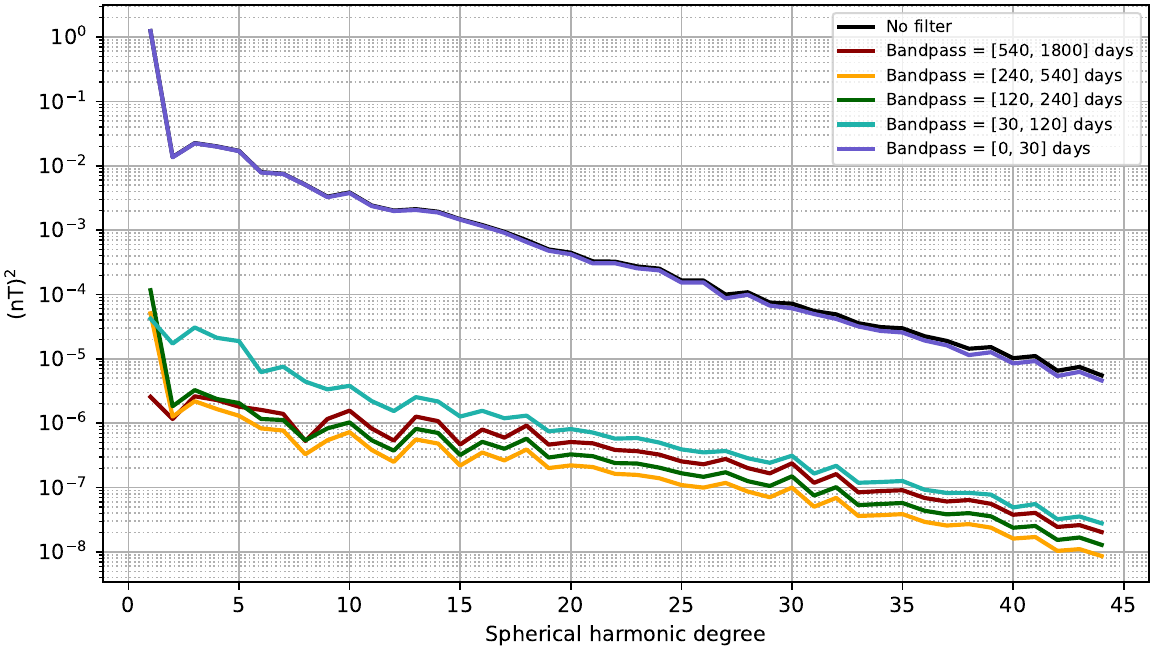}\hspace{0.1cm}\includegraphics[width=3.1in]{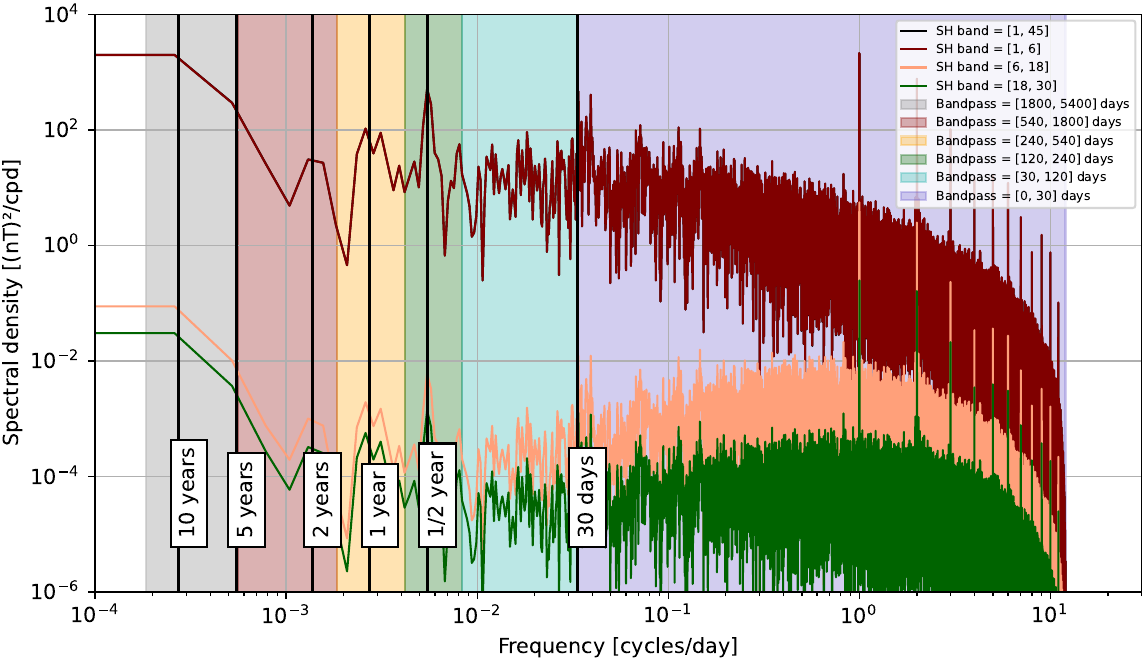}
\caption{Left: Spatial spectra of the mean squared vector magnetic field at satellite altitude (450\,km) associated with Earth-induced internal currents produced by magnetospheric variations from the CHAOS-7 magnetospheric field model\cite{Finlay:2020} acting on a 3D ocean+1D mantle conductivity model, calculated using the  method of Grayver \cite{Grayver2021b} and averaged over 2014.0 to 2024.0.  As well as the unfiltered signal (black curve), spatial spectra are shown for interannual, annual, semi-annual, monthly and sub-monthly period bands (coloured curves). Right: Temporal power spectral density $\hat{P}_{B_r}$ at satellite altitude of the radial component of the same Earth-induced internal field, unfiltered (black) and in chosen spherical harmonic bands (curves can overlap). These curves are the median of power spectral densities, computed using a Welch-type method from time series of $B_r$ computed on an approximately equal area grid of 3000 points at satellite altitude.}
\label{fig:magneto_spectra}
\end{figure}

Finally we turn to the field induced in the conducting Earth by variations of magnetospheric currents.  Here we do not explicitly consider the magnetospheric field itself, since we assume it can be adequately separated from the internal field by potential field modelling. To characterize the magnetospheric induced field we have used the CHAOS-7 \cite{Finlay:2020} magnetospheric field model, that includes a parameterization of the near-Earth magnetospheric ring current with time-dependence given by the ground-based RC index, and a simple treatment of fields due to magnetopause and magnetotail currents \cite{Olsen:2014} capturing daily and annual variations related to changes in geocentric-solar-magnetic coordinates with respect to the Earth-fixed-Earth-Centered frame \cite{Laundal:2017}.  This was used as the inducing field and the induced response of the conducting Earth was computed in the time domain using the method of Grayver \cite{Grayver2021b} again using a 3D ocean and sediment electrical conductivity model above a 1D electrically conducting mantle \cite{Grayver2017}.  Spatial and temporal spectra of the resulting internal induced field are presented in Fig.\ref{fig:magneto_spectra}.

The spatial spectra of the internal Earth-induced field produced by magnetospheric variations are dominated by large length scales and most of the power occurs on periods less than 30 days.  The temporal power spectral density does nevertheless show power across all frequencies though without the very strong spectral lines found for the ionospheric field (the weak lines in Fig.\,\ref{fig:magneto_spectra} on daily and high harmonics are artifacts due to a restriction to night-side ground observatories applied when computing the RC index).  There are nevertheless noticeable spectral peaks at annual and especially semi-annual periods that are likely related to variations of the Earth's orbital plane with respect to the solar wind and associated variations in magnetospheric current systems as seen on the Earth \cite{Russell1973,Malin1976,Malin1999}. The power present at annual and semi-annual periods in the magnetospheric variations (and also the ionospheric variations presented previously) is clearly much larger than the OIMF signals at these periods.  Methods for separating out large external scale fields (for example resulting in the green and orange lines in Fig.\,\ref{fig:magneto_spectra}, right) are clearly essential if the OIMF field is to be extracted.

As for the ionospheric field these results have been computed from a global internal (induced) field each hour, without selecting geomagnetically quiet times.  It therefore includes geomagnetic storms for which significant induced fields are expected.  Focusing on geomagnetic quiet times a weaker magnetospheric induced signal is expected, we return to this question in the next section. 

\section{Strategies for extracting the OIMF}
\label{sect:strategies}
Based on the above characteristics of the various geomagnetic field sources at satellite altitude we now discuss possible strategies for extracting the OIMF signal.

\begin{figure}[!h]
\centering\includegraphics[width=5in]{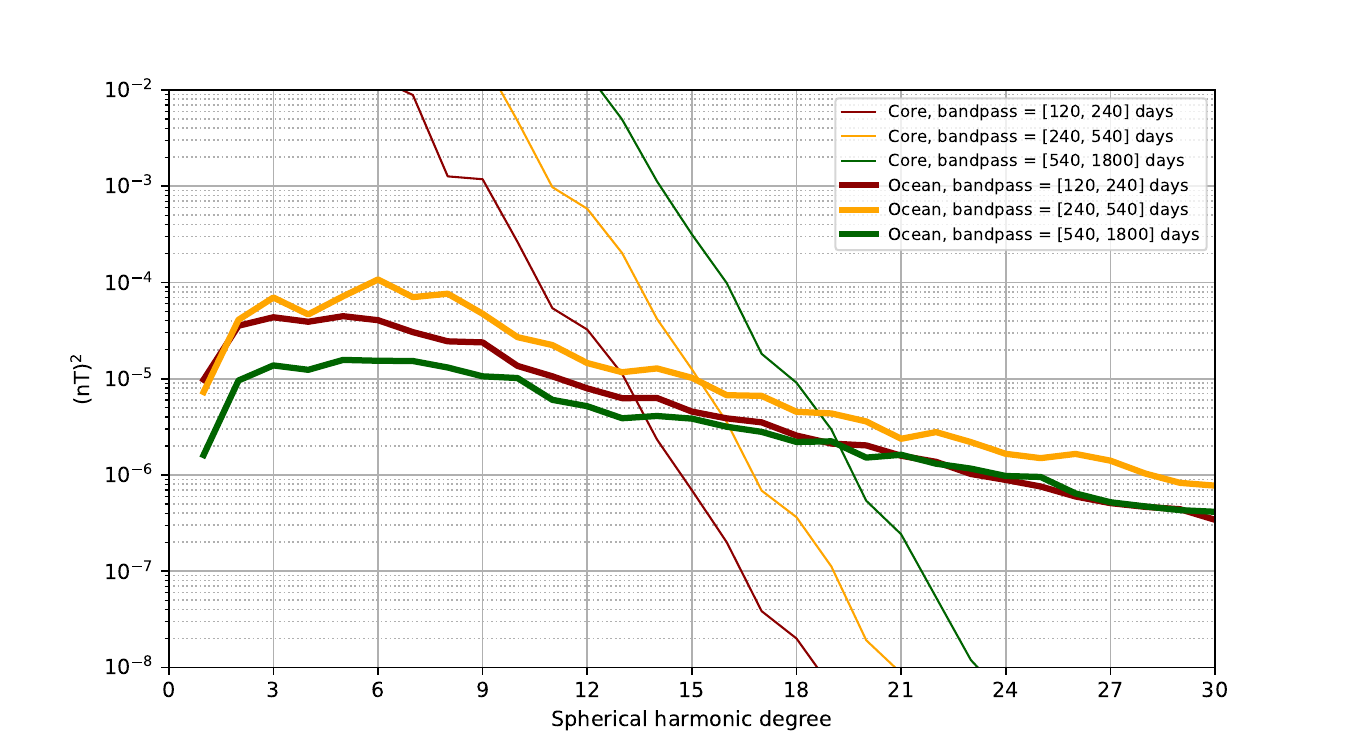}
\caption{Comparison of spatial spectra of the oceanic induced and core magnetic fields, at satellite altitude (450\,km), up to spherical harmonic degree 30.  The spectra cross, so that the ocean signal dominates above degree 13 to 19 depending on the period considered.}
\label{fig:crossing_spectra}
\end{figure}

Given the dominant role of the core field in the near-Earth magnetic field, an obvious first question is whether the OIMF is always obscured below the core field.  In Fig. \ref{fig:crossing_spectra} we present comparisons of the spatial power spectra for the core and ocean induced signals at satellite altitude, considering the semi-annual (120-240 day), annual (240-540 day) and interannual (540-1800 day) period bands.  We find the OIMF spectra cross the core fields spectra at degrees 13 to 19; the crossing takes place at lower degree for shorter periods.  This suggests that we should expect the OIMF to dominate over the core field signal on short length scales and short time scales, for example at spherical harmonic degrees greater than 15 for annual periods.  Recall that considering the annual period band the signal due to the nearly steady crustal field should also be negligible.

\begin{figure}[!h]
\centerline{Inversion from synthetic data, after filtering \qquad \qquad}
\centering\includegraphics[width=5.0in]{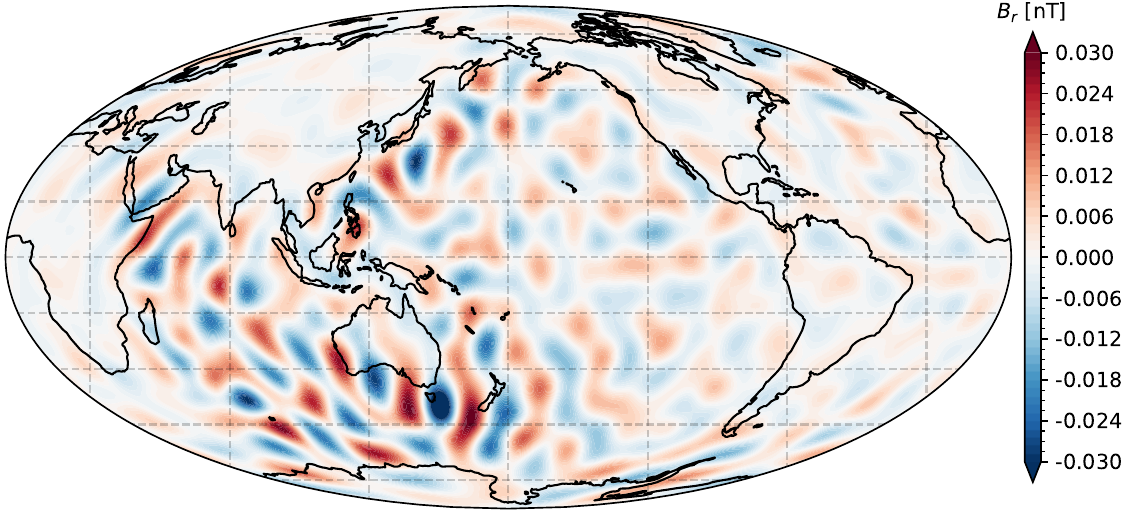}
\centerline{Reference Simulation, after filtering \qquad }
\centering\includegraphics[width=5.0in]{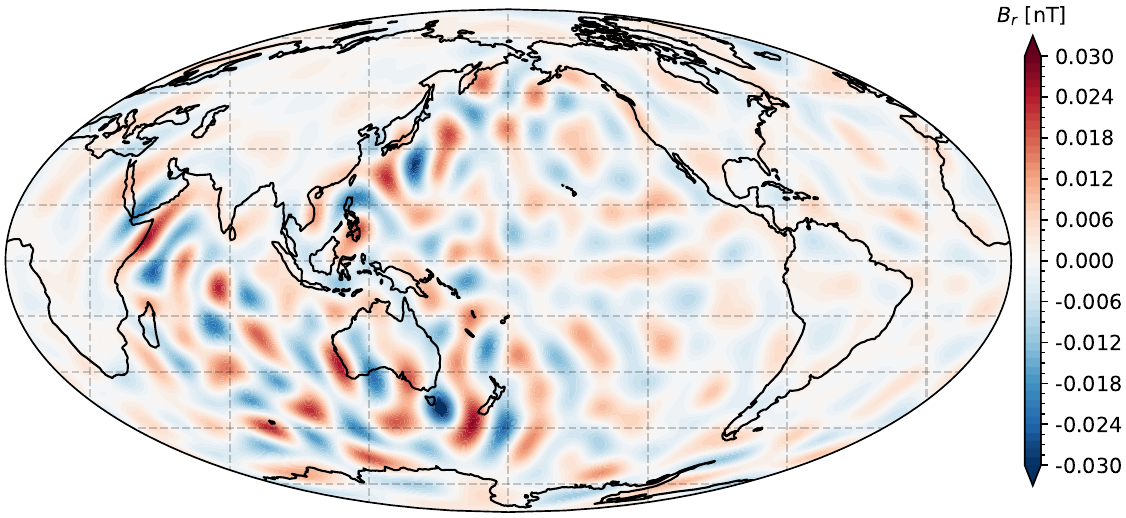}
\caption{Map of the radial component of the ocean-induced magnetic field at satellite altitude (450\,km), filtered to retain a period band centered on 1 year (passband 240 to 540 days) and spherical harmonic degrees 16 to 30. Top: result from inversion of synthetic satellite data, Bottom: For reference, simulated OIMF used to generate the synthetic satellite data.}
\label{fig:Br_ocean_surf_1yband}
\end{figure}

To further test this hypothesis, we carried out a field modelling experiment, evaluating our simulated core field and OIMF along the actual orbits of the \textit{Swarm} A, B and C satellites between 2014.0 and 2024.0.  We applied geomagnetic quiet-time and dark selection criteria, and used vector and vector gradient data at non-polar latitudes, scalar intensity data in the polar region, and scalar gradients at all latitudes, similar to the CHAOS-7 model \cite{Finlay:2020}.  We inverted for a time-varying internal field up to spherical harmonic degree 30 and parameterized time-dependence using cubic splines with a 0.15 year knot spacing.  The resulting radial field at satellite altitude, filtered to show the signal in the 1 year period band, and only spherical harmonic degrees 16 to 30, is presented in Fig.\ref{fig:Br_ocean_surf_1yband}, along with the input OIMF signal for reference.  We are clearly able to successfully retrieve the OIMF signal from degree 16 to 30 in this one year period band, despite the presence of a realistic core field. Physically interesting signals are evident especially in the equatorial Pacific (likely associated with westward wave propagation) and in the Indian ocean (related to monsoon-driven seasonal variations in the ocean circulation). 

\begin{figure}[!h]
\centering\includegraphics[width=3.1in]{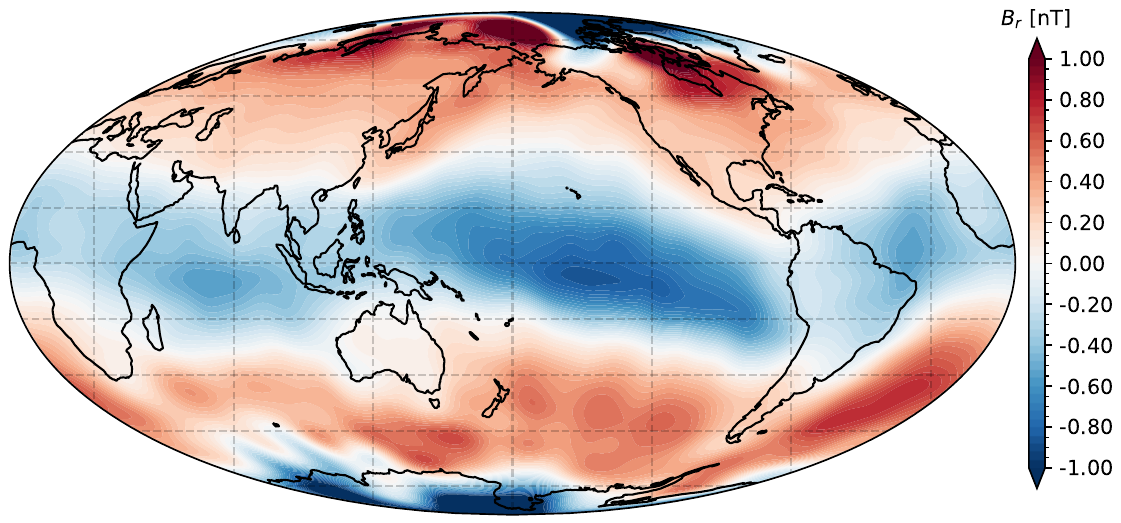}\hspace{0.05cm}\includegraphics[width=3.1in]{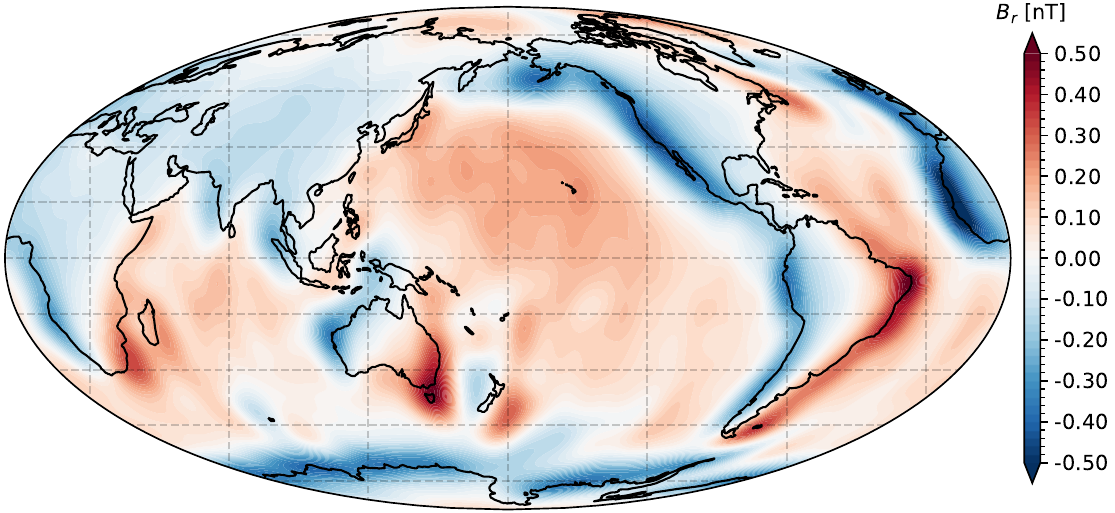}
\caption{Left: Ionospheric and associated induced field in the annual period band, estimated by inversion of synthetic satellite data from dark/night regions, based on the CI version 9 model \cite{Sabaka:2018}.  Radial field is shown at 450km altitude, up to spherical harmonic degree 30 in 2016.5. Right: Magnetospheric induced internal field in the annual period band, at 450km altitude, up to spherical harmonic degree 30 in 2016.5, based on the CHAOS magnetospheric model \cite{Finlay:2020} acting on a 3D ocean conductivity and sediments model with an underlying 1D conducting Earth \cite{Grayver2021b}.}
\label{fig:mag_iono_maps}
\end{figure}

This synthetic test is however incomplete as we have ignored the additional fields related to ionospheric and magnetospheric sources. Fig.\,\ref{fig:mag_iono_maps} shows results of similar inversions carried out based on synthetic satellite data containing only the ionospheric and associated induced signals (left panel), and only the magnetospheric induced signals (right panel).  Unlike the results shown in the earlier sections, these inversion results now take into account the dark and geomagnetic quiet-time conditions that are important for the assessing the impact of magnetospheric and ionospheric signals when estimating internal field signals.  

Both the magnetospheric induced signal and the ionospheric and associated induced signals estimated in these inversions are much larger than the OIMF signal in the annual period band, even when considering dark and geomagnetically quiet conditions.  However both are rather large scale, at least outside the polar region (for the ionospheric field) and away from the ocean continent boundaries (the magnetospheric induced field).  A simple approach to isolating the OIMF might therefore involve spatially filtering to remove these magnetospheric and ionospheric components.  For example by excluding spherical harmonic degrees below 16 (already necessary to exclude the core field) and possibly also the low order (near zonal) part of the field.  Such crude filtering will, of course, also remove OIMF signals of interest and may not even be sufficient to remove all the ionospheric and magnetospheric fields.  A better approach will certainly be to make use of ionospheric and magnetospheric field models, such as those we have used here for our synthetic tests, removing their predictions from the satellite data before they are used in field modelling.

\section{Discussion and conclusions}
Using models for the different contributions to Earth's magnetic field we have shown that the long period ocean induced magnetic field is measurable at satellite altitude and that it dominates over the core field signal provided one considers sufficiently short length and time scales. 

The general circulation OIMF field signal at such length and time scales, for example at spherical harmonic degree above 15 in the annual period band, is however expected to be rather small on the basis of the models considered here, on the order of 0.02\,nT. The most serious obstacles to retrieving this signal are ionospheric and associated Earth-induced fields, and Earth-induced fields driven by magnetospheric field variations, both of which, like the OIMF, are internal to satellite orbits.  Our tests suggest these are associated with rms signals of order of 0.15\,nT and 0.05\,nT respectively at satellite altitude, for spherical harmonic degrees 16-30 in the annual period band; these means 90\% and 50\% respectively of these signals would need to be corrected or filtered out in order for the OIMF to be retrieved.

Progress towards meaningful satellite magnetic monitoring of the OIMF will require that these confounding signals are sufficiently well modelled and removed from satellite data before inverting for the OIMF. It is not the day-to-day variability of the magnetospheric and ionospheric signals that it is crucial here, rather their slow variations on the monthly to decadal timescales of interest for the OIMF. Inversion strategies such as regularization at Earth's surface, or use of prior model covariances based on the expected fields from advanced OIMF simulations could also help.  The application of filters during post-processing that are designed to remove remaining signal of ionospheric and magnetospheric origin will also likely be essential.  

The geomagnetic models we have employed here involve some known limitations.  Polar ionospheric fields are not fully described in the CI model and we have effectively ignored field-aligned currents.  Our characterizations will therefore underestimate the complexity of real geomagnetic data in the polar region.  For this reason we recommend that investigations of the OIMF should focus, at least to begin with, on signals at mid and low latitudes.  For example, our simulations suggest the existence of locally strong signals in the Indian Ocean in the annual period band.  The numbers quoted above for the typical OIMF and magnetospheric/ionospheric field amplitudes are global rms values; it might be significantly easier to extract the OIMF in particular locations.  There are also limitations in our knowledge of the mantle electrical conductivity, in particular in its deep 3D structure.  Depending on the true mantle conductivity the magnetospheric induced field could be more complex than that we have investigated here, however deep mantle structures are expected to primarily impact the observed field on large lengthscales.  It should also be acknowledged that the investigations presented above were based on a single advanced OIMF simulation, there are uncertainties in the ocean flows and the conductivity structures that serve as its inputs.  The true OIMF could therefore also differ from that presented here, although experience with the tidal OIMF suggests simulations of the type considered here should be fairly reliable guides \cite{Grayver2024}.

Despite the challenges described above we believe that detecting the OIMF due to ocean dynamics is now feasible and worth a concerted effort by the geomagnetic community.  Magnetic remote sensing of ocean dynamics would provide valuable independent constraints on ocean dynamics, and has the potential to eventually contribute, along with other observations, to our knowledge of changes in ocean heat content and deep salinity variations.

The past decade with continuous magnetic field observations from the three \textit{Swarm} satellites have seen important advances in our modelling of ionospheric and magnetospheric fields.  These will be further improved in the upcoming years thanks to the data from the MSS-1 satellite and its sucesssors, and the planned NanoMagSat satellites, which will provide much improved local time coverage from space.  With \textit{Swarm} also descending closer to the ocean sources during the next solar minimum there could be a bright future ahead for magnetic remote sensing of ocean dynamics, but it still remains to be demonstrated with real observations that the long period OIMF signal can be reliably extracted.  

\vskip6pt

\section*{Acknowledgements}
This work has been funded by ESA Contract No. 4000143627/24/I-EB  Swarm for Ocean Dynamics, part of the ESA Solid Earth Magnetic Science Cluster, EO for Society programme.  The presented simulations of the ocean induced magnetic field were supported by the Ministry of Education, Youth and Sports of the Czech Republic through the e-INFRA CZ (ID:90254), project
OPEN-30-29. CF thanks Alexander Grayver for valuable discussions, sharing Q matrix kernels for a 3D+1D Earth conductivity model and assistance in computing the induced counterparts of the  magnetospheric field. Sasha Troncy-Portier is thanked for providing initial versions of the power spectral density and temporal filtering functions.


\bibliographystyle{RS}
\bibliography{refs}

\end{document}